\documentclass[prd,onecolumn,preprint,nofootinbib]{revtex4}
\usepackage[plainpages=false, colorlinks=true, anchorcolor=blue, linkcolor=blue, citecolor=blue, bookmarks=false, urlcolor=red]{hyperref}
\usepackage[utf8]{inputenc}
\usepackage{color}

\usepackage{float}
\usepackage{amsfonts,amsmath,amssymb}
\usepackage{natbib}
\usepackage{array}
\usepackage{textcomp}
\usepackage{gensymb}
\usepackage{subcaption}
\usepackage{caption}
\usepackage{booktabs}
\usepackage{lipsum} 
\usepackage{lineno} 

\newcommand{\rthis}[1]{\textcolor{black}{#1}}

\usepackage{graphicx}
\graphicspath{{./images/}}

\begin{document}

\title{Search for transient gamma-ray emission from magnetar flares using Fermi-LAT}

\author{Vyaas \surname{Ramakrishnan}}
\altaffiliation{Mail: vyaas3305@gmail.com}
\author{Shantanu \surname{Desai}}
\altaffiliation{Mail: shntn05@gmail.com}

\begin{abstract}
        We search for transient gamma-ray emission in the energy range from 0.1-10 GeV using data from the Fermi-LAT telescope in coincidence with magnetar flares.
        For our analysis, we look for coincidence with 15 distinct flares from 11 magnetars using three time windows of $\pm$ 1 day, $\pm$ 7 days,  and $\pm$ 15 days.
        For 14 of these flares from 10 magnetars, we do not see any statistically significant gamma-ray emission.
        However, we see one possible gamma-ray flare from one magnetar, namely 1E 1048.1-5937, with combined significance of   $4.4\sigma$, (after considering the look-elsewhere effect) observed after about 10 days from the start of the X-ray flare.
        However, this magnetar is located close to the galactic plane (with galactic latitude of -0.52\degree), and this signal could be caused by contamination due to diffuse flux from gamma-ray sources in the galactic plane. 
\end{abstract}

\affiliation{Department of Physics, IIT Hyderabad, Kandi, Telangana-502284, India}

\maketitle

\section{Introduction}
    Magnetars are a distinct class of neutron stars distinguished by their extraordinarily strong magnetic fields, which can reach extremely high values  up to $10^{15}$G and X-ray luminosities between $10^{31}-10^{34}$ ergs/sec~\cite{Woods,Mereghetti,Kaspirev,Espositorev,magnetars25}.
    The magnetar emission stems from the decay and instability associated with their ultra-strong magnetic fields~\cite{DuncanThompson,Paczynski,ThompsonDuncan}.
    Observationally, they are associated with soft-gamma ray repeaters (SGRs) and anomalous X-ray pulsars (AXP)~\cite{Mereghetti}.
    Recently, a very intense radio burst characteristic of fast radio bursts, has also been detected from a galactic magnetar (SGR 1935+2154), which is a soft gamma-ray repeater~\cite{Chime}, hinting towards an association between magnetars and FRBs. 
    Magnetars are known to emit a diversity of X-ray transients including short bursts, large outbursts with energies between $10^{40}-10^{46}$ ergs, giant flares, bursts with quasi-periodic oscillations, etc.~\cite{Kaspi14}.
    These outbursts are emitted on a wide range of time scales from a few seconds to years~\cite{Kaspi14,Espositorev}.
    In this work, we focus on searching for gamma-ray emission from magnetars in coincidence with these X-ray outbursts. 
    
    Some theoretical models of gamma-ray emission from magnetars have previously been discussed in Refs.~\cite{Cheng01,Zhang02,Takata13}.
    The first ever theoretical calculation of the high energy gamma-ray flux for seven AXPs was done in ~\cite{Cheng01}.
    This work showed that the gamma-ray radiation arises due to curvature radiation produced by acceleration of electron-positron pairs, which are produced from collisions of high energy photons in the outer gap and soft X-rays from the stellar surface.
    This work argued that gamma-rays produced by magnetars could be detected by Fermi-LAT. Subsequently, the gamma-ray flux during the quiescent magnetar phase was calculated for two SGRs, viz. SGR 1806-20 and SGR 1900+14~\cite{Zhang02}, and shown to be detectable for SGR 1900+14 by Fermi-LAT~\cite{Zhang02}.
    A model of pulsed GeV gamma-ray emission from magnetars at the time of X-ray outbursts was also proposed in \cite{Takata13}.
    In this model, the X-ray outbursts were proposed to be triggered by crust fractures, which could excite the Alfven waves that propagate into the outer magnetosphere.
    The GeV gamma rays are then produced by curvature radiation at the outer magnetosphere~\cite{Takata13}.
    However, no flux predictions for individual magnetars were made in this work. 
    As discussed in ~\cite{Fermi10}, even null results stemming from gamma-ray emission from magnetars provide interesting constraints on curvature/synchrotron emission mechanism in magnetars at high latitudes. 
    
    In this work, we search for gamma-ray emission using the Fermi Large Area Telescope (LAT). The LAT is one of the two instruments onboard this detector and is sensitive to high-energy gamma rays from a plethora of astrophysical sources~\citep{Ajello19}. This detector is a pair-conversion telescope with a field of view of about 2.4 sr that is sensitive to photons between the energy range of 30 MeV to around 300 GeV and has been operating continuously since 2008~\citep{Atwood09}. 
    
    Previous searches for gamma-ray emission from magnetars with Fermi-LAT have been carried out in ~\cite{Fermi10,Li17}. The first work which looked for gamma-ray emission from 0.1-10 GeV, was done by the Fermi-LAT collaboration using 17 months of Fermi-LAT data from 13 magnetars~\cite{Fermi10}. No statistically significant emission was detected from any of the magnetars. Stringent upper limits on the persistent emission between $\sim 10^{-12}$ and $10^{-10}$ erg $\rm{s^{-1} cm^{-2}}$ were obtained~\cite{Fermi10}. Then, a follow-up search using about six years of Fermi-LAT data was carried out in ~\cite{Li17}, which looked for gamma-ray emission in the range from 0.1-10 GeV from 20 magnetars. No significant emission from these magnetars was detected and stringent upper limits between $10^{-12}$ and $10^{-11}$ erg $\rm{s^{-1}~cm^{-2}}$ were obtained~\cite{Li17}. In addition to galactic magnetars, the Fermi-LAT collaboration also found evidence for GeV emission from an extragalactic magnetar in the Sculptor galaxy~\cite{fermilatmagnetardet}. Therefore, with more than 12 years of data, a search for gamma-ray emission in coincidence with magnetar outbursts using the latest data is timely, since the previous search for gamma-ray emission was only done in 2017.

    For this work, we follow the same methodology as our previous works on analysis using Fermi-LAT, which looked for gamma-ray emission from galaxy clusters and OJ 287~\cite{MannaFermi,OJA}. This paper is structured as follows. We introduce the use of {\tt easyFermi}, a software tool that largely simplifies the analysis of Fermi LAT data in Sect.~\ref{sec:analysis}. We then  into the individual analysis of magnetar outbursts using dynamic time ranges and observe the obtained spectral energy distributions and light curves in Sect.~\ref{sec:results}. We conclude in Sect.~\ref{sec:conclusions}.

\section{Observations and Data Analysis}
\label{sec:analysis}
    For this work, we have obtained the time ranges of magnetar bursts observed in X-rays from the Magnetar Outburst Online Catalog\footnote{Magnetar Outburst Online Catalog : http://magnetars.ice.csic.es/\#/outbursts}~\cite{Esposito}. The peak X-ray luminosity during these outbursts is between $1-5 \times 10^{35}$ ergs/sec. We have looked for a gamma-ray signal using 15 individual outbursts from 11 different magnetars. Among these 11 magnetars, 7 are SGRs and 4 AXPs. The McGill magnetar catalog~\cite{Kaspi14}, on the  other hand, lists 30 known magnetars. However, for the remaining 19 magnetars, the outbursts (if any) were detected before the launch of the Fermi-LAT satellite in 2008, and hence were not considered in this work.
    
    The durations of the magnetar flares are listed in Table~\ref{tab:mag_data}.
    We also note that the observed peak of the magnetar flare (as in MOOC) is almost always coincident with the onset of the burst, based on the first available datum in the X-ray light curve, with the exceptions of SGR 0418+5729, SGR 1833-0832, and 1E 1048.1-5937 (corresponding to the flare at 57592 MJD), where the observed peak of the flare occurred 8 days, 5 days, and 6 days, respectively, after the onset.
    

    The analysis of the Fermi-LAT data is done using {\tt easyFermi}~\cite{R.Menezes}, which largely simplifies the data collection for analysis. The data files involving the spacecraft and photon files can either be downloaded directly from Fermi-LAT Data Query or can be automatically fetched using {\tt easyFermi}, by specifying the magnetar's J2000 coordinates and the time duration. This analysis uses data from the 3rd data release of LAT data\footnote{4FGL-DR3: https://fermi.gsfc.nasa.gov/ssc/data/access/lat/12yr\_catalog/}, which covers an extended 12 year time range~\cite{DR3,4FGLpart2}. The Pass 8 Source v3 instrument response functions (IRFs) were used in the analysis. {\tt easyFermi} also provides an option for downloading the required galactic (gll\_iem\_v07.fits) and isotropic (iso\_P8R3\_SOURCE\_V3\_v1.txt) diffuse emission models.
    
    For this work, we consider gamma-ray photons within the standard energy range of 0.1-10 GeV and a circular region of interest (RoI) of 10.6\degree. To minimize the gamma-ray contamination from Earth's limb, we adopt the standard zenith angle cut of 90\degree. The maximum likelihood analysis was performed on a $15\degree \times 15\degree$ square region with a spatial bin size of $0.1\degree$ and eight energy bins per decade, similar to \cite{Fermi10,Li17}. The spectral-spatial model includes all the cataloged sources lying in the square region $10\degree$ larger than the extent of the $15\degree \times 15\degree$ region. The magnetars are included in this model as point sources with a power-law spectrum, using their positions as listed in the McGill Online Magnetar Catalog. Source fitting is done using \emph{fit} from {\tt fermipy} with \emph{NewMinuit} as the optimizer. All spectral parameters are left free for sources within $5\degree$ of the RoI center, while the spectral parameters of the other sources are left to their catalog values. The magnetars in Table \ref{tab:mag_data} are modeled with a power law spectrum with all spectral parameters allowed to vary. In all analyses, we ensure a fit quality of 3, which is indicative of an accurate fit. 

    The analysis, carried out by {\tt easyFermi}, involves a binned likelihood analysis using the \emph{GTAnalysis} class in {\tt fermipy}~\cite{fermipy}.
    For this purpose, a maximum likelihood test statistic, \textit{TS} is used to evaluate the credibility of the gamma-ray source i.e. to quantify the likelihood of a source being real. The test statistic is defined as twice the maximum log-likelihood difference between an alternate and null hypothesis of no signal, or rather, $TS = 2\ln({L_{max, 1}/L_{max, 0})}$~\cite{Mattox}, where $L_{max, 1}$ would signify the maximum likelihood value for a model with an additional test source (the `alternate' hypothesis) and $L_{max, 0}$ signifies the maximum likelihood value for a model without an additional source (the `null' hypothesis). The same statistic is also used in neutrino astronomy~\cite{Pasumarti} and soft gamma-ray astronomy~\cite{MannaMeV}. A higher TS value for a spatial bin indicates that the alternate hypothesis is valid, or in simpler terms, it is more likely that the observed gamma-ray flux in that spatial bin is due to a legitimate source.
    For a source consisting of four free parameters (two for position and two for flux and spectral index), the distribution of TS is close to 1/2 of $\chi^2$ distribution with four degrees of freedom~\cite{Abdo10}. Therefore in this case a TS of 25 would correspond to a $p$-value of $2.5 \times 10^{-5}$ or 4.1$\sigma$. For our analysis,  since the two coordinates of the source are known, we have two degrees of freedom. Therefore, a TS value of 25, then corresponds to a statistical significance of $\sim 4.6 \sigma$~\cite{Abdo10}. 
    
    \begin{table}[H]
    \tiny
        \centering
        \captionsetup{justification=centering}
        \begin{tabular}{|c|c|c|c|c|c|c|c|}
            \hline
            \textbf{Source} & \textbf{Flare Start} & \textbf{Flare Start Date} & \textbf{Outburst} & \textbf{TS} & \textbf{TS} & \textbf{0.1 - 10 GeV} & \textbf{4FGL srcs}\\
            & \textbf{(MJD)} & \textbf{(Calendar)} & \textbf{Timescale} & \textbf{($\pm  15$ days)} & \textbf{($\pm 1$  day)} & \boldmath$(\Gamma = 2.5)$ & \textbf{within 5\degree}\\
            \hline
            Swift J1822.3-1606 & 55758 & 2011-Jul-16 & 219 days & 0.00 & 6.45 & $<\:2.34$ & 56\\
            Swift J1834.9-0846 & 55780 & 2011-Aug-07 & 97 days & 0.00 & 0.00 & $<\:2.95$ & 51\\
            1E 1048.1-5937* & 55926 & 2011-Dec-31 & 949 days & 24.91 & 0.00 & $-$ & 45\\
            1E 1048.1-5937* & 57592 & 2016-Jul-23 & 211 days & 2.51 & 0.00 & $<\:3.98$ & 45\\
            1E 1048.1-5937* & 58120 & 2018-Jan-02 & 140 days & 0.00 & 0.00 & $<\:4.63$ & 45\\
            1E 1547.0-5408* & 54742 & 2008-Oct-03 & 23 days & 0.00 & 0.00 & $<\:2.87$ & 29\\
            1E 1547.0-5408* & 54854 & 2009-Jan-23 & 640 days & 0.00 & 0.00 & $<\:2.38$ & 29\\
            1E 2259+586* & 56045 & 2012-Apr-28 & 1298 days & 4.21 & 0.00 & $<\:0.95$ & 27\\
            CXOU J1647-4552* & 55829 & 2011-Sep-25 & 28 days & 1.27 & 2.37 & $<\:4.32$ & 77\\
            SGR 1935+2154 & 57075 & 2015-Feb-22 & 302 days & 0.00 & 0.00 & $<\:1.65$ & 21\\
            SGR 1935+2154 & 57526 & 2016-May-18 & 71 days & 0.00 & 0.00 & $<\:1.75$ & 21\\
            SGR 0418+5729 & 54993 & 2009-Jun-11 & 637 days & 0.00 & 0.00 & $<\:1.01$ & 9\\
            SGR 1833-0832 & 55275 & 2010-Mar-20 & 159 days & 0.00 & 0.00 & $<\:3.97$ & 52\\
            SGR 0501+4516 & 54701 & 2008-Aug-23 & 373 days & 3.48 & 0.00 & $<\:1.33$ & 6\\
            SGR 1745-2900 & 56411 & 2013-Apr-29 & 1177 days & 8.30 & 0.00 & $<\:4.60$ & 116\\
            \hline
        \end{tabular}
        \caption{Properties of magnetars studied in this work, including the start times of magnetar outbursts (as obtained by the first datapoint in the MOOC-based light curve), sorted in order of magnetar along with the start date of the flares (MJD). We note that the peak of the X-ray flare coincides with the onset of the flare based on the available data for all magnetar outbursts, except for SGR 0418+5729, SGR 1833-0832 and 1E 1048.1-5937 (corresponding to the flare at 57592 MJD) for which the offset is eight, five, and six days, respectively. We have also shown the duration of the flare starting from the onset to quiescence obtained using data available from the MOOC website~\cite{Esposito} for all outbursts except for the flare in 1E 1048.1-5937 starting at 58120 MJD, for which we obtain the duration from ~\cite{Kaspi20}. The energy flux upper limits (0.1 - 10 GeV), at 95\% confidence level, provided above are the ones obtained through the analysis of a $\pm$15 days time window, with the spectral index of the magnetar point sources fixed at 2.5. Fluxes are in the units of $10^{-11}\:\rm{erg\:cm^{-2}\: s^{-1}}$. The magnetars marked with an asterisk(*) are AXPs (Anomalous X-Ray Pulsars), while the rest are SGRs (Soft Gamma Repeaters).}
        \label{tab:mag_data}
    \end{table}

    The results of the {\tt easyFermi} analyzes also include spectral energy distributions (SEDs) and light curves. The former is computed using the \emph{sed} method from {\tt fermipy}, which fits the flux normalizations of the target source in each individual energy bin. In our analyses, we have used 12 energy bins to plot the SEDs. In specific cases, though, we use fewer energy bins to better observe a weaker signal. The light curves of the magnetar point sources are generated using the \emph{lightcurve} method from {\tt fermipy}, which fits the source characteristics through a series of time bins. We have adopted a standard value of 15 time bins for the light curves. For each magnetar in Table \ref{tab:mag_data}, we present the results for  the analysis in two distinct time windows ($\pm$1 day and $\pm$15 days), centered around the onset of the magnetar outburst. Similar to ~\cite{fermilatmagnetardet}, we also did the analysis in $\pm$ 7 day window, but since no additional signal was seen, these results are not reported. In the following sections, we shall go through the transient analysis results for each of the magnetars in detail.

\section{Results}
\label{sec:results}
    We now present our results for gamma-ray emission during the magnetar outbursts. For almost all the magnetars, we find that there is no conclusive evidence for gamma-ray emission coinciding with the onset of the magnetar outburst. For these magnetars, the TS values were much less than 25. Any signals in the vicinity of the magnetar were due to sources in the 4FGL-DR3 point source catalog~\cite{DR3,4FGLpart2}. Consequently, in such cases, we report the energy flux upper limits obtained through the likelihood analysis. The 95\% confidence level energy flux upper limits (0.1 - 10 GeV energy range) for magnetars, are then calculated using a power-law spectrum, with a fixed photon index of 2.5, as in~\cite{Fermi10} and~\cite{Li17}.
    
    The only exception was for the magnetar 1E 1048-5937, wherein we observe a TS value $\approx 25$ for the first outburst at MJD of 55926, which has a significance of $4.4\sigma$, based on a one month time window and incorporating the look-elsewhere effect, considering the fact that we have considered three distinct temporal windows. We now discuss the results for 1E 1048-5937 followed by the other magnetars. A tabular summary of all our results  can be found in Table~\ref{tab:mag_data}.
    
    \subsection{Results for  1E 1048.1-5937 around MJD of 55926}
        1E 1048.1-5937 is an Anomalous X-ray Pulsar (AXP), at around 9 kiloparsecs away.
        A long-lasting X-ray flare was detected by SWIFT-XRT between 0.5-10 keV, beginning at 55926 MJD and having a peak flux of $\sim 3 \times 10^{-11}$ $\rm{erg/sec/cm^2}$ and a decay time of 550 days~\cite{Archibald15,Kaspi20}.
        The Fermi-LAT analysis of 1E 1048.1-5937's outburst at 55926 MJD yielded a high TS value of 24.91 over a 1-month time window.
        However, the TS value drops to zero during the 2 day window centered on the magnetar X-ray flare.
        This suggests that the gamma-ray emission during the larger, one month period is not exactly coincident with the peak in the X-ray emission.
                
\begin{figure}[H]
            \centering
            \includegraphics[width=0.49\linewidth]{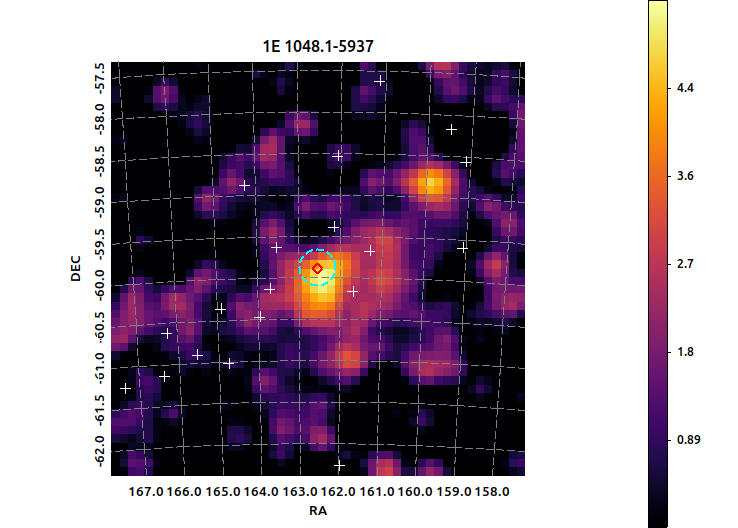}
            \includegraphics[width=0.49\linewidth]{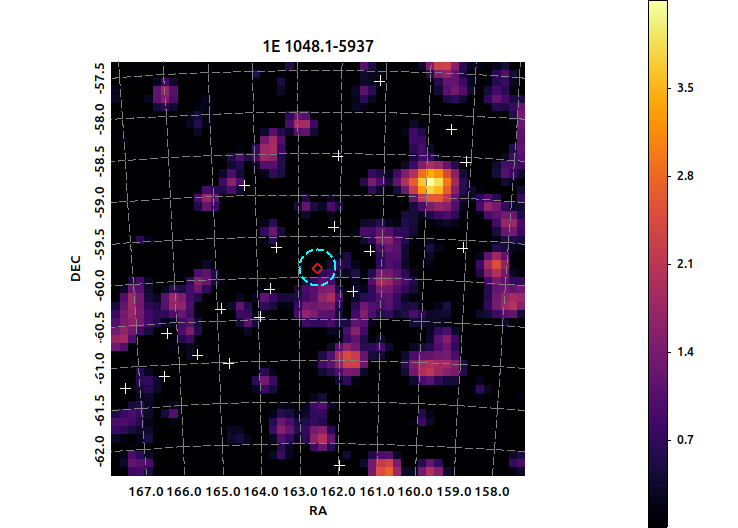}
            \caption{\small{The Fermi-LAT fields for 1E 1048.1-5937 corresponding to its first burst (MJD 55926). In the left plot, the detection significance obtained from  $\sqrt{TS}$ map for the 0.1-10 GeV energy range are illustrated. The plot on the right highlights the residual $\sqrt{TS}$ maps, obtained after modeling the $\gamma$-ray emission from the magnetar. Both maps are based on a 1-month analysis ($\pm 15$ days) window centered on MJD 55926. \rthis{The red diamond indicates the position of the magnetar point source used in the analysis, inferred from the McGill magnetar catalog~\cite{Kaspi14}}. The cyan dotted circle denotes the 99\% positional uncertainty radius of the localized magnetar point source.}}
            \label{fig:1e-plots}
        \end{figure}

        The results of the analyses can be summarized efficiently using a spatial significance map.
        The maps cover a $5\degree \times 5\degree$ square region with a spatial bin size of $0.1\degree$.
        These significance maps can be found  in Fig.~\ref{fig:1e-plots}.
        In Fig.~\ref{fig:1e-plots}, we show two significance maps.
        The left panel presents the significance map derived by excluding 1E 1048.1-5937 from the model, effectively highlighting the $\gamma$-ray emission attributable to the magnetar.
        In contrast, the right panel depicts the residual significance map obtained after incorporating the magnetar into the region of interest (RoI) model, thereby representing the residual $\gamma$-ray emission after accounting for the contribution from 1E 1048.1-5937.
        There is a very visible peak in the significance maps, (with significance of  $4.4\sigma$), coinciding with the magnetar's first X-ray outburst at 55926 MJD.
        
        Upon performing a spatial localization for the magnetar point source, the best-fit position is given by (ra:162.424$^{\circ}$, dec:-60.000$^{\circ}$), with a 99\% positional uncertainty radius of 0.22$^{\circ}$. We also note that the maxima in TS lies within this region, as illustrated in Fig.~\ref{fig:1e-plots}.
        
        Additionally, we tested the incorporation of a new unidentified gamma-ray source exactly coinciding positionally with the maxima in TS, along with the magnetar point source. On performing a spatial localization for the unidentified source, we determine that the best-fit position is (ra:162.412$^{\circ}$, dec:-60.013$^{\circ}$), with a 99\% positional uncertainty radius of 0.22$^{\circ}$. Notably, 1E 1048.1-5937 lies well within this region and furthermore, no existing Fermi-LAT sources are found within this region in the 4FGL-DR3 or 4FGL-DR4 LAT catalogs~\cite{DR3,DR4} within the 99\% positional uncertainty. 

        We also looked for other possible multi-wavelength candidate counterparts, using {\tt SIMBAD}, within the 99\% positional uncertainty radius of 0.22$^{\circ}$, which corresponds to a 13.2 arc-minute radius. We find a total of 345 sources, including one high-mass X-ray binary (2E-2336) at an angular separation of 13.58 arc-seconds and five X-ray point sources with angular separations ranging from $\approx 330 - 645$ arc-seconds, from the magnetar. Other sources include main-sequence stars, eclipsing binary stars, long period variable stars and RR Lyrae variable stars. \rthis{These sources (e.g. the high-mass X-ray binary 2E 2336) might be viable alternatives to the source of the observed gamma-ray emission.}

        Further, we modeled the magnetar as an extended source using a radial disk model.
        The results indicate that there is no significant statistical evidence supporting the emission being attributed to an extended source, as the extension hypothesis test statistic, $TS_{ext} = 0.0002$ is negligible.
        
\begin{figure}[H]
            \centering
            \includegraphics[width=0.49\linewidth]{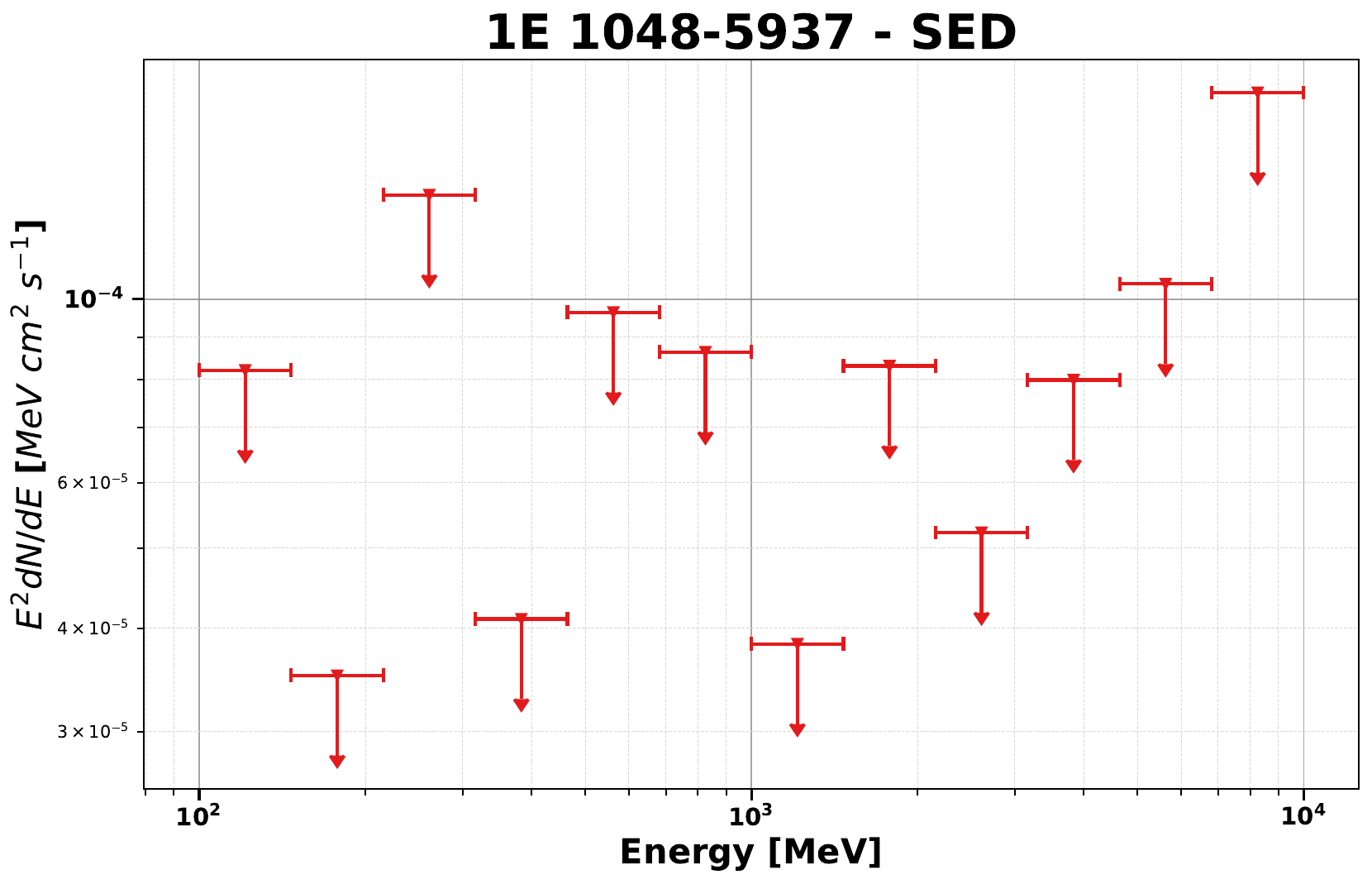}
            \includegraphics[width=0.49\linewidth]{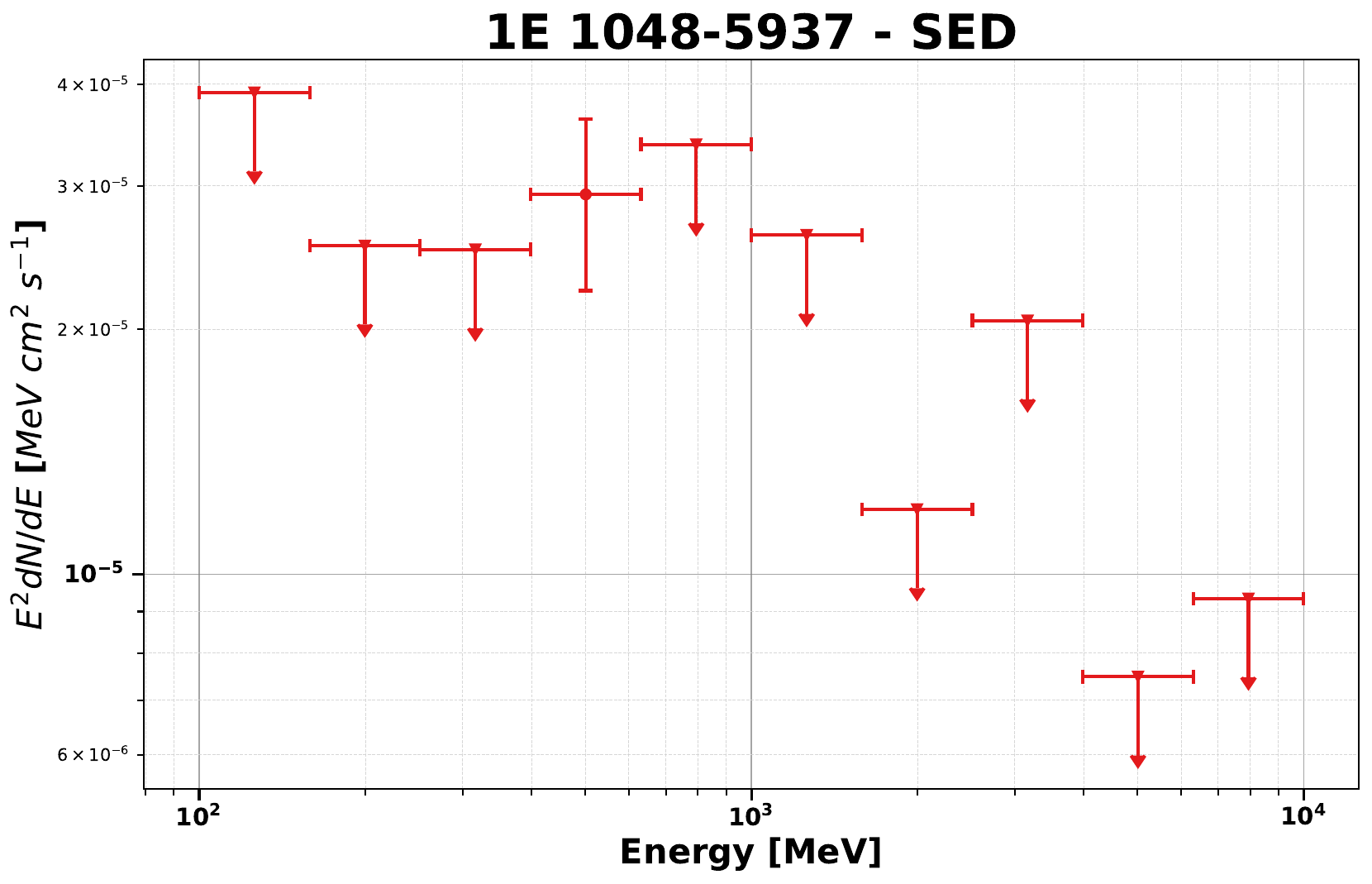}
            \caption{\small{Spectral Energy Distribution (SED) plots for 1E 1048.1-5937. The left panel presents the SED derived from a 2-day observational window, while the right panel displays the SED from the analysis of a 1-month time window, both centered around 55926 MJD}}
            \label{fig:sed-1e1048}
        \end{figure}

        The spectral energy distributions (SED) for the first outburst are presented in Fig.~\ref{fig:sed-1e1048}. The SED, obtained from the analysis of the one month time period, shows a detected flux point in the energy bin at 500 MeV.

\begin{figure}[H]
            \centering
            \includegraphics[width=0.85\linewidth]{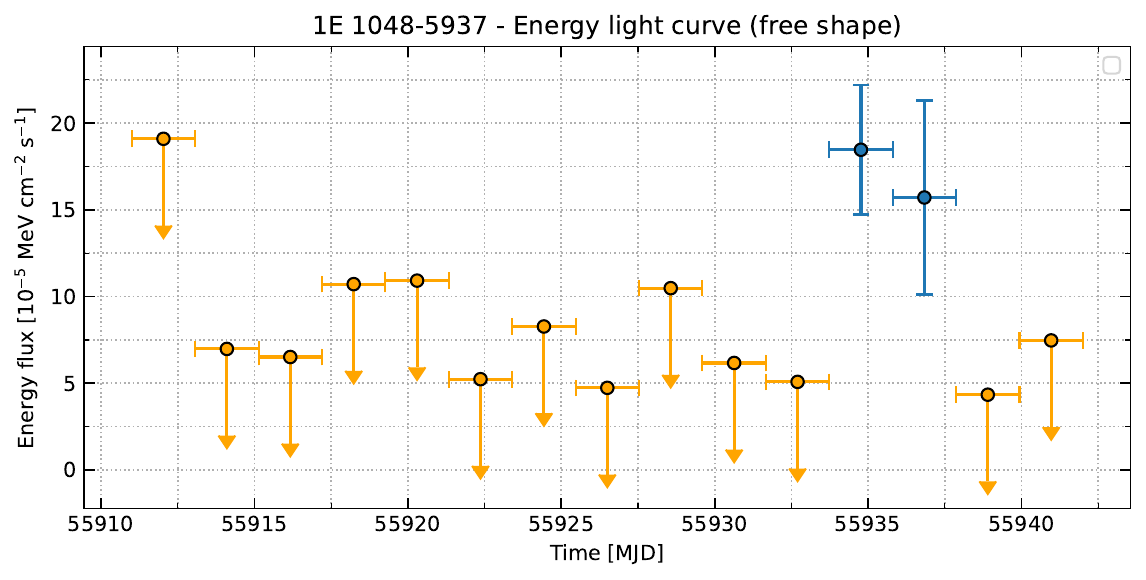}
            \caption{\small{Energy flux curve of 1E 1048.1-5937, derived from a 1-month time window, using the standard energy range of 0.1 - 10 GeV.}}
            \label{fig:mlc-std-1e1048}
        \end{figure}

        We also look at the standard $\gamma$-ray light curve to get a better idea of the duration of the gamma-ray signal. We adopt a standard binning of 15 bins, over the 1-month time range. The $\gamma$-ray light curve is provided in Fig.~\ref{fig:mlc-std-1e1048} for the 1-month time range. In these light curves, all data points with $TS<9$ are shown as upper limits, whereas if $TS\geq 9$, the corresponding flux along with its Poisson error is shown.

\begin{figure}[H]
            \centering
            \includegraphics[width=\linewidth]{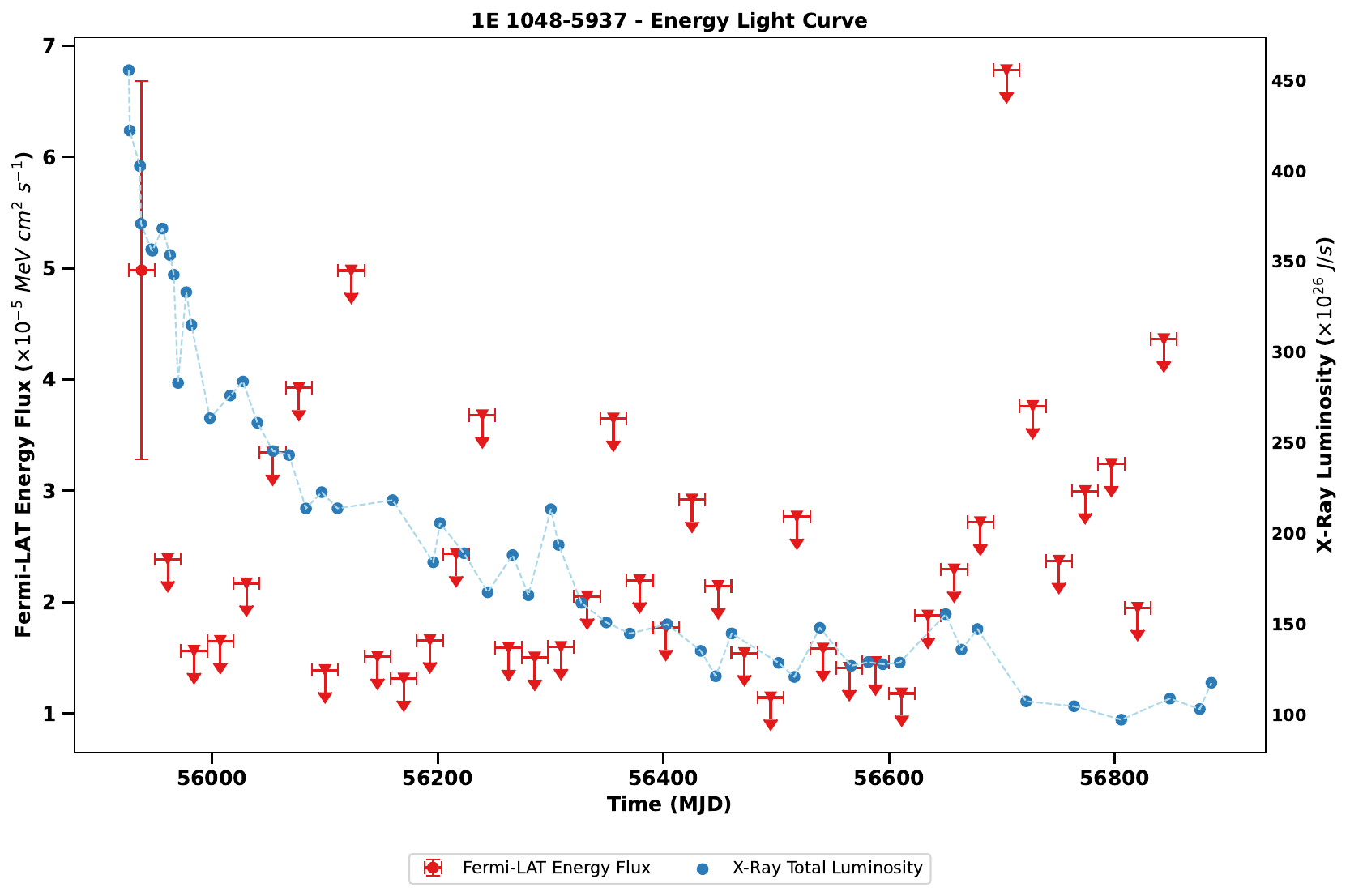}
            \caption{\small{Comparison of the Fermi LAT energy flux (in the range 0.1-300 GeV) along with the  total X-Ray luminosity of 1E 1048.1-5937, over a longer time span from 31/12/2011 to 17/07/2014}. The X-ray luminosity data is from the MOOC website.}
            \label{fig:dual-1e1048}
        \end{figure}

        Further, we overlay the X-ray and Fermi-LAT $\gamma$-ray light curves for 1E 1048.1-5937.
        In order to directly compare the gamma-ray energy light curves from our Fermi LAT analysis to the X-Ray luminosity curves available on MOOC\footnote{Magnetar Outburst Online Catalog : http://magnetars.ice.csic.es/\#/outbursts}, we plot both the X-ray and gamma-ray light curves in the same figure, to check if there is a direct correlation between the two. This plot showing both the  fluxes can be found in Fig.~\ref{fig:dual-1e1048}.
        Therefore, with only one data point, it is hard to make a definitive statement about the correlation between the X-ray and gamma-ray light curve.
        We also note that the TS value of the first data point (with a bin-size of 23.2 days) in Fig.~\ref{fig:dual-1e1048} is about 15.65 and is suppressed, compared to that obtained within a one month time window.

\begin{figure}[H]
            \centering
            \includegraphics[width=0.49\linewidth]{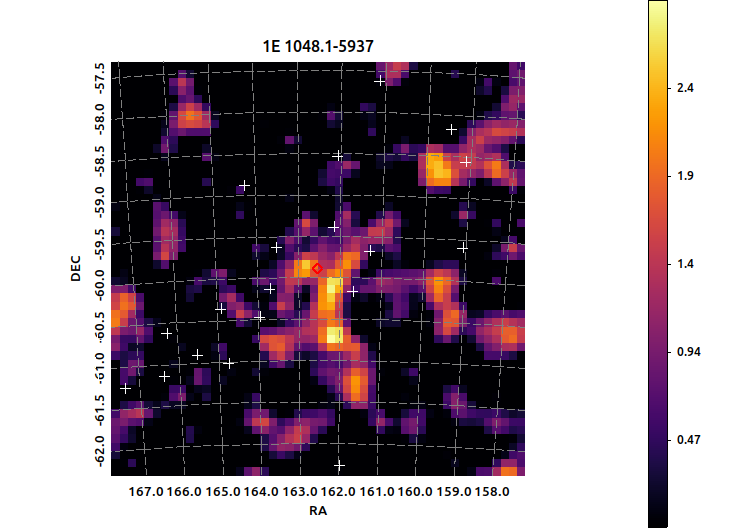}
            \includegraphics[width=0.49\linewidth]{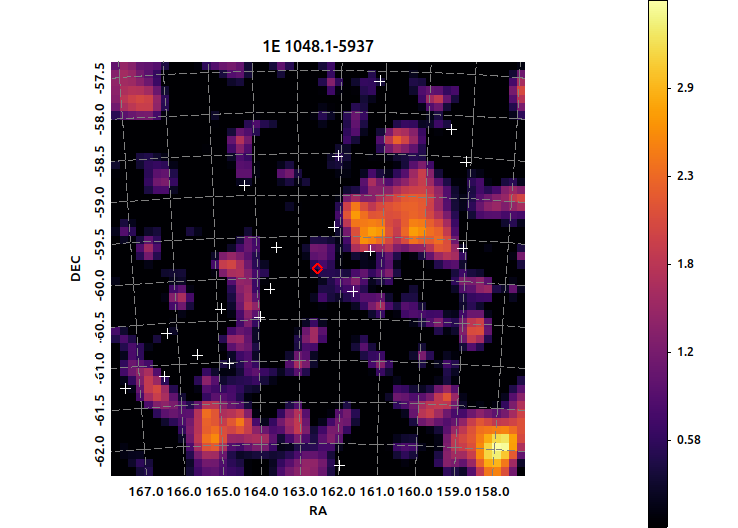}
            \caption{\small{The significance maps (given by $\sqrt{TS}$) for the Fermi-LAT field for 1E 1048.1-5937 corresponding to its second burst (MJD 57592) is shown in the left plot. The Fermi-LAT field corresponding to its third burst (MJD 58120) is shown in the right plot}}
            \label{fig:1e-b2b3}
        \end{figure}

        In addition to the observed burst at MJD 55926, two other X-ray flares were detected for this magnetar at MJD of 57592 and 58120.
        The timescales of these two flares are also provided in Table \ref{tab:mag_data}. The duration of the flare corresponding to 1E 1048.1-5937's third outburst in early 2018 is obtained from ~\cite{Kaspi20}.
        However, no gamma-ray emission was detected corresponding to these two flares.
        The spatial significance maps for the latter two, smaller X-ray outbursts at MJD of 57592 and 58120 are provided in Fig.~\ref{fig:1e-b2b3}.
        We can clearly see that the TS value is much less than 9 and no gamma-ray signal is detected in coincidence with these flares.
        
    Therefore, for the first X-ray outburst at 55926, we find a gamma-ray signal from 1E 1048.1-5937 about 10 days after the peak of the X-ray flare. 
    No gamma-ray signal is seen corresponding to the other two X-ray outbursts for this magnetar.
    Finally, we caution that this magnetar is located close to the galactic plane where diffuse emission is strong and could contaminate the observed signal~\cite{Fermi10}. 
    
\subsection{Results from other magnetars}
\begin{figure}[H]
        
            \centering
            \includegraphics[width=\linewidth]{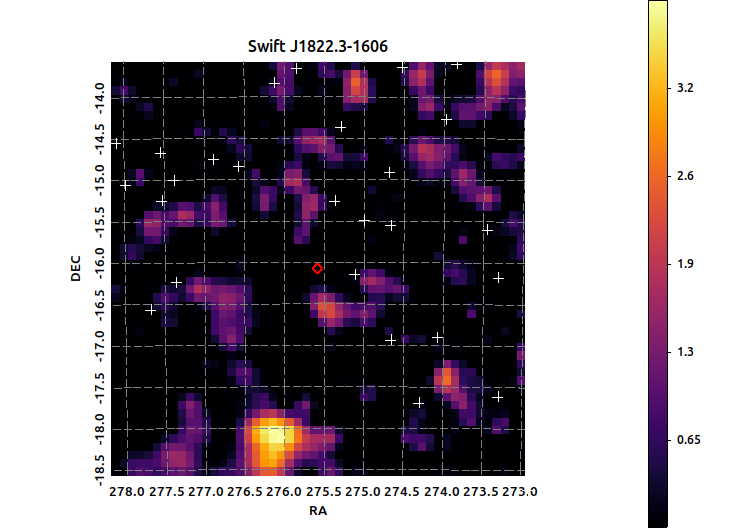}
            \caption{Significance map (given by $\sqrt{TS}$) for Swift J1822.3-1606.  The maps span a $5\degree \times 5\degree$ region with a spatial bin size of $0.1\degree$. The white crosses indicate 4FGL-DR3 point sources and the red diamond denotes the magnetar point source.
             }
            \label{1822}
        \end{figure}

\begin{figure}[H]
            \centering
            \includegraphics[width=\linewidth]{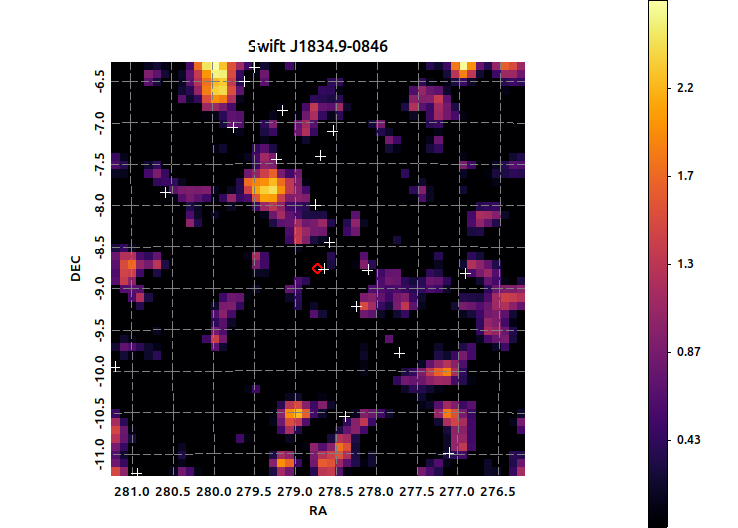}
            \caption{Significance map  for  Swift J1834.9-0846 similar to Fig.~\ref{1822}. The nearest 4FGL-DR3 associated source overlapping the X-ray position of the magnetar  is 4FGL J1834.5-0846e (spp). }
            \label{1833}
     \end{figure}

\begin{figure}[H]
                 \centering
            \includegraphics[width=0.49\linewidth]{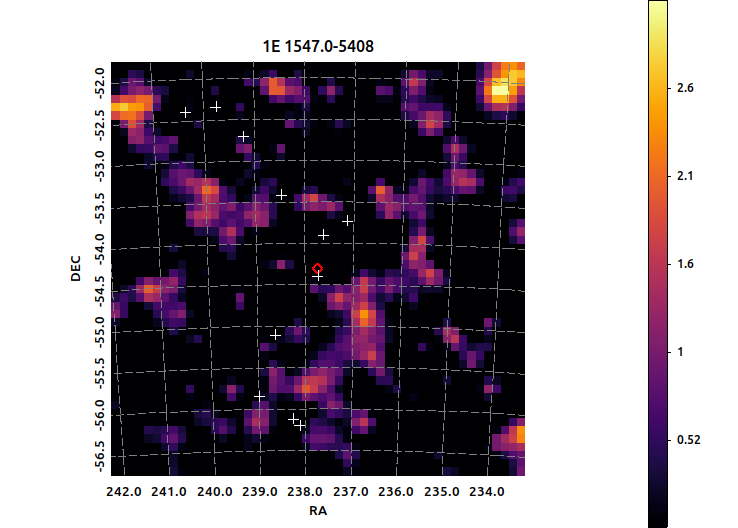}
            \includegraphics[width=0.49\linewidth]{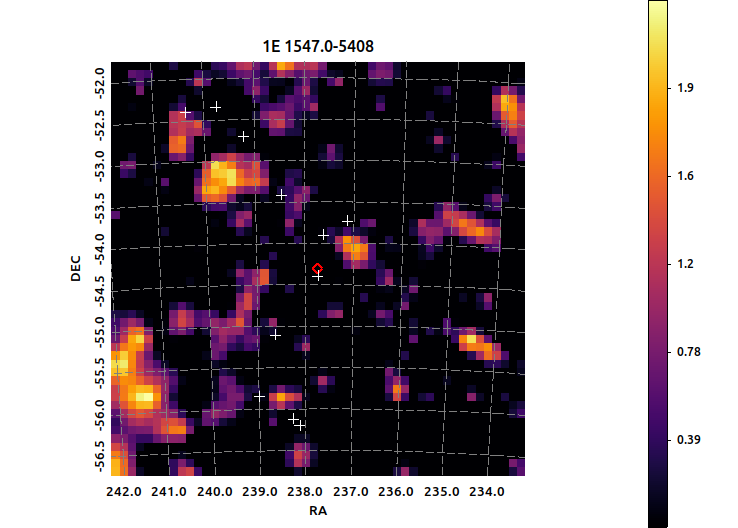}
            \caption{Significance map (similar to Fig.~\ref{1822}) for  1E 1547.0-5408 corresponding to its first outburst at MJD 54742 (left panel) and second outburst at MJD of 54854  (right panel). The nearest 4FGL-DR3 associated source is 4FGL J1550.8-5424c (spp).}
            \label{1547}
 \end{figure}

\begin{figure}[H]
                 \centering
            \includegraphics[width=\linewidth]{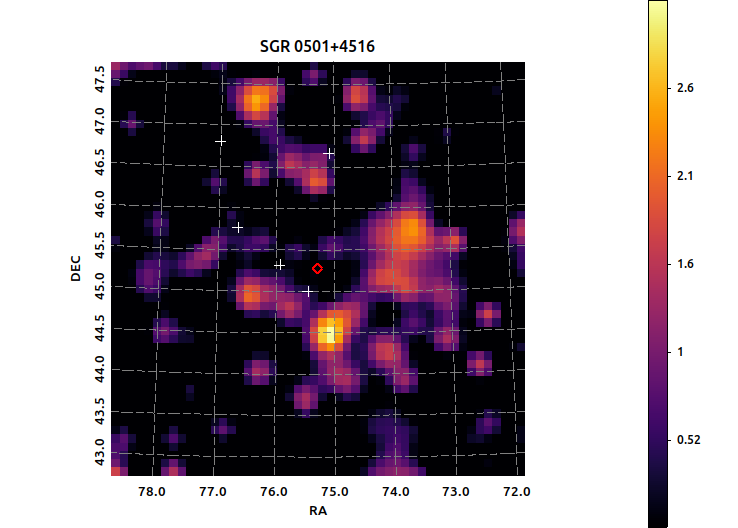}
            \caption{Significance map (similar to Fig.~\ref{1822}) for SGR 0501+4516.}
             \label{SGR0501}
        \end{figure}

\begin{figure}[H]
            \centering
            \includegraphics[width=\linewidth]{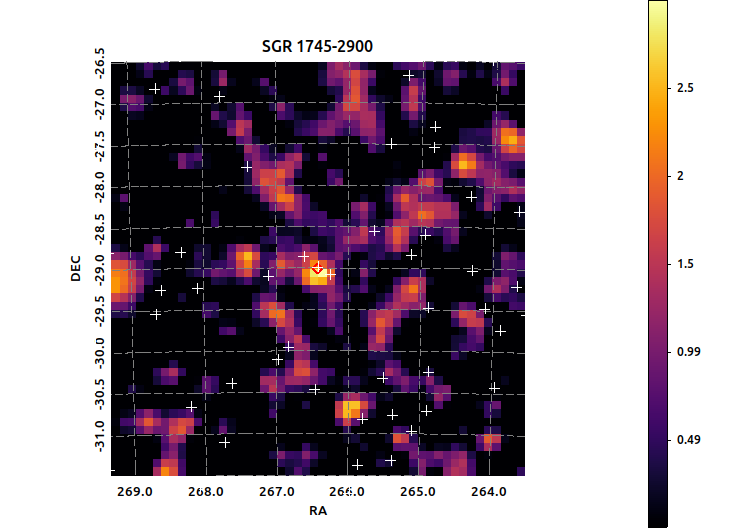}
            \caption{Significance map (similar to Fig.~\ref{1822}) for  SGR 1745-2900. The nearest 4FGL-DR3 associated source is 4FGL J1745.6-2859 (Galactic center)~\cite{Fabio}.}
            \label{SGR1745}
    \end{figure}

\begin{figure}[h]
               \centering
            \includegraphics[width=\linewidth]{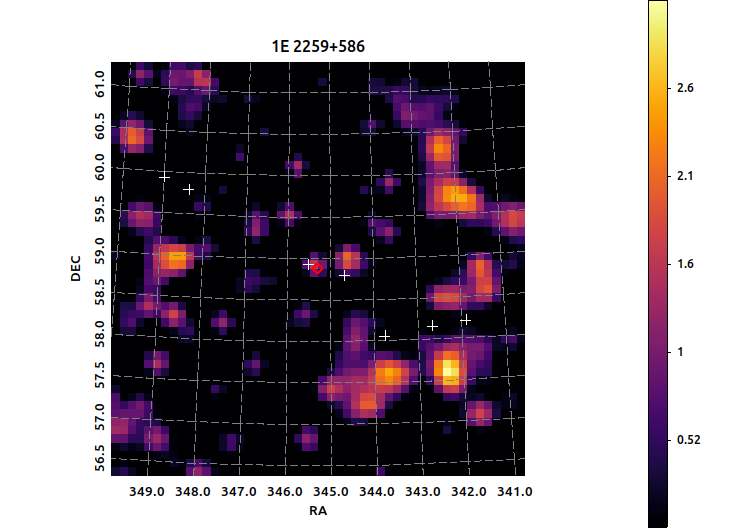}
            \caption{Significance map (similar to Fig.~\ref{1822}) for   1E 2259+586. The nearest 4FGL-DR3 associated source is 4FGL J2301.9+5855e (SNR).}
            \label{1E2259}
        \end{figure}

\begin{figure}[H]
            \centering
            \includegraphics[width=\linewidth]{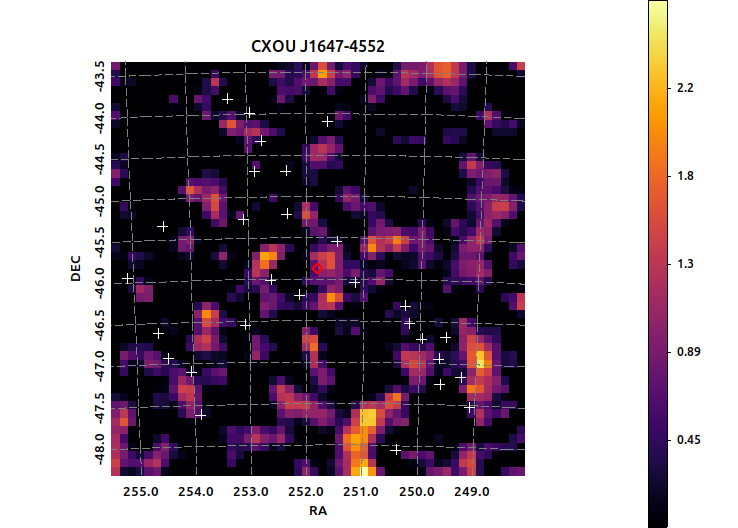}
            \caption{Significance map (given by $\sqrt{TS}$) for  CXOU J1647-4552.}
            \label{CXOU}

    \end{figure}

\begin{figure}[H]
            \centering
            \includegraphics[width=0.49\linewidth]{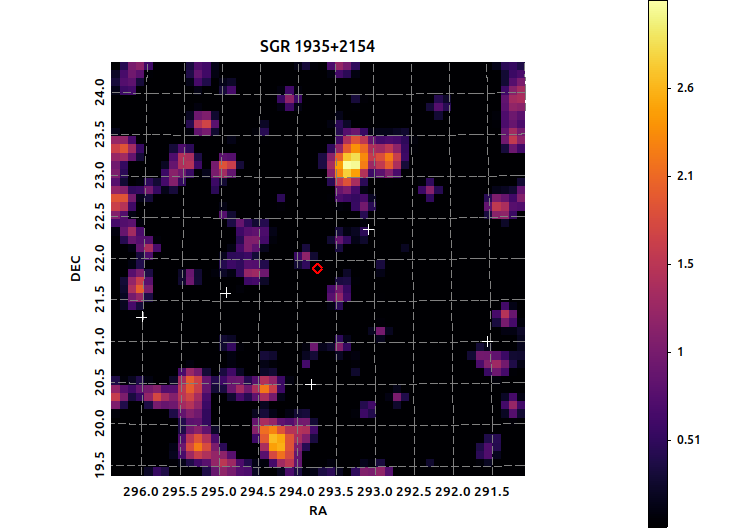}
            \includegraphics[width=0.49\linewidth]{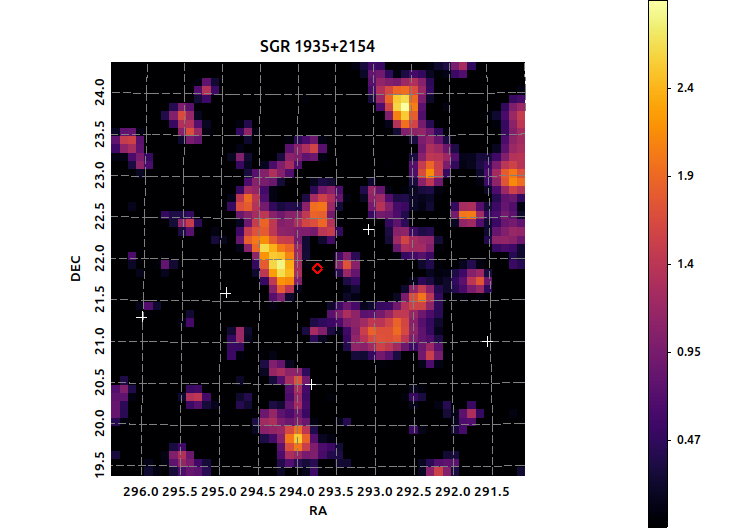}
            \caption{Significance map (similar to Fig.~\ref{1822}) for SGR 1935+2154 corresponding to its first burst at MJD 57075 (left panel) and second burst at MJD 57526  (right panel).}
            \label{SGR1935}
        \end{figure}

    We now present our results for searches for gamma-ray signals for all other magnetars around the time of X-ray detected flares presented in Table~\ref{tab:mag_data}.
    For our analysis, we choose a one-month window around the X-ray flare observed from the magnetar.
    One of the remaining magnetars includes SGR 1745-2900 at the Galactic center~\cite{Fabio}, which was excluded in previous searches due to difficulty in resolving the source from the surrounding environments~\cite{Li17}.
    A few magnetars such as SGR 1935+2154 and 1E 1547.0-5408 emitted multiple bursts.
    We looked for a gamma-ray signal in a one-month window around all  these bursts.
    The spatial significance maps (obtained from $\sqrt{TS}$) covering a square spatial region of $5\degree \times 5\degree$, can be found in Fig.~\ref{1822}, Fig.~\ref{1833}, Fig.~\ref{SGR0501}, Fig.~\ref{SGR1745}, Fig.~\ref{1E2259}, and Fig.~\ref{CXOU} for Swift J1822.3-1606, Swift J1834.9-0846, SGR 0501+4516, SGR 1745-2900, 1E 2259+586, and CXOU 1647-4552, respectively.
    The corresponding spatial maps for both the outbursts of SGR 1935+2154 can be found in Figs.~\ref{SGR1935}, whereas those for 1E 1547.0-5408 can be found in Fig.~\ref{1547}.
    
    Some of the magnetars are close to  Fermi-LAT 4FGL-DR3 point sources, which are in near proximity to the magnetars.
    These include Swift J1834.9-0846, 1E 1547.0-5408, SGR 1745-2900, 1E 2259+586 and SGR 1833-0832.
    These are shown using a white-plus marker, while the X-ray positions of the magnetars are depicted using a red diamond.
    The spectral spatial model used in the analyses of these sources also accounts for the $\gamma$-ray emission from these Fermi-LAT 4FGL-DR3 point sources.
    The closest Fermi-DR3 point sources are 4FGL J1834.5-0846e (spp)\footnote{``spp'' stands for ``SNR (Supernova remnant), Pulsar or PWN (Pulsar wind nebula)'', and refers to sources of unknown nature but overlapping with known SNRs or PWNe and thus candidate members of these classes~\cite{4FGLpart2}}, associated with Swift J1834; 4FGL J1550.8-5424c (spp), associated with 1E 1547; 4FGL J1745.6-2859 (Galactic center), associated with SGR 1745-2900; 4FGL J2301.9+5855e (Supernova remnant), associated with 1E 2259; and 4FGL J1834.2-0827c*, associated with SGR 1833.
    The source classes of the nearest sources are given in parentheses, while the unclassified sources are denoted by an asterisk.
    More details about these nearby 4FGL-DR3 point sources and source classes can be found in ~\cite{4FGLpart2}.
    Apart from these known sources, no other statistically significant $\gamma$-ray signal with significance $> 5\sigma$ is detected within a one month span of the X-ray flares for each of these magnetars.
    The 95\% confidence level energy flux limits (0.1 - 10 GeV), for all of the magnetars can be found in Table~\ref{tab:mag_data}.
    Finally, for some of the magnetars for which the X-ray flares last for greater than 600 days, we also did an extended search for a gamma-ray signal in a +200 days window starting from the peak of the flare. However, we do not detect any signal (with $TS>10$) in such an extended window.
    
    
    
    \begin{figure}[H]
            \centering
            \includegraphics[width=\linewidth]{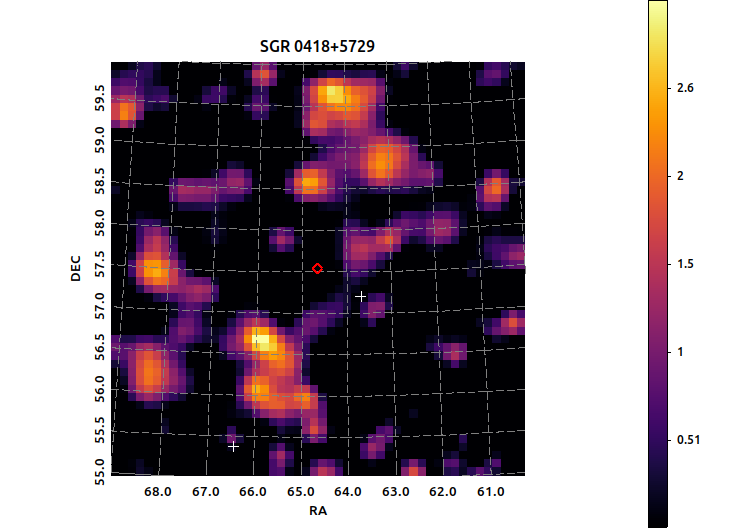}
            \caption{Significance map (similar to Fig.~\ref{1822}) for  SGR 0418+5729.}
           \label{SGR0418} 
        \end{figure}
        \begin{figure}[H]
            \centering
            \includegraphics[width=\linewidth]{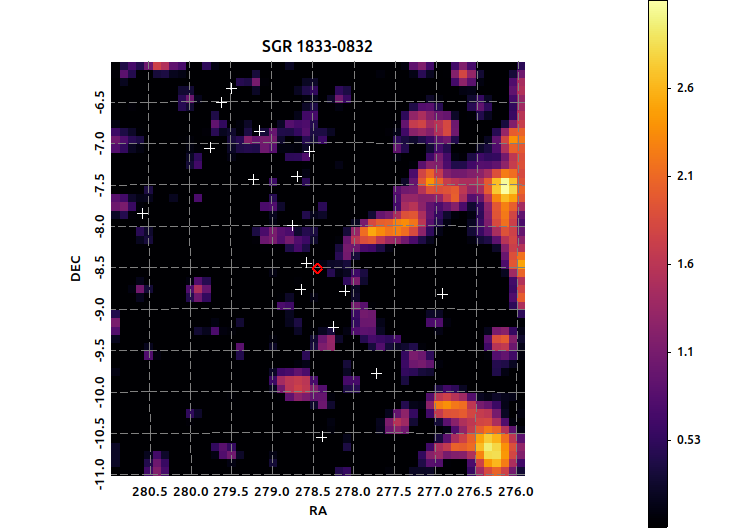}
            \caption{Significance map (similar to Fig.~\ref{1822}) for  SGR 1833-0832. The nearest 4FGL-DR3 associated source is 4FGL J1834.2-0827c*.}
            \label{SGR1833} 
        \end{figure}

\section{Conclusions}
\label{sec:conclusions}
    In this work, we have searched for transient gamma-ray emission between 0.1-10 GeV from magnetar flares using 12 years of Fermi-LAT data.
    For this purpose, we searched in three temporal windows of $\pm$ 1 day, $\pm$ 7 days,  and $\pm$ 15 days around 15 distinct flares from 11 different magnetars.
    The times of magnetar outbursts were obtained using X-ray data based on the compilation in ~\cite{Esposito}. 
    We do not detect any gamma-ray emission above $3\sigma$ in coincidence with almost all the flares.
    The corresponding 95\%  upper limits on the energy flux for these can be found in Table~\ref{tab:mag_data} and are between $\sim 1-5 \times 10^{-11} \rm{erg~cm^{-2}~s^{-1}}$. We note that the upper limits are of the same order of magnitude as those obtained in ~\cite{Abdo10,Li17}.
    The only magnetar for which we see gamma-ray emission (with significance of 4.4$\sigma$) is for 1E 1048.1-5937, for which the gamma-ray signal is seen around 10 days after the peak of the X-ray flare with observed energy flux of around $(18.0 \pm 3.7) \times 10^{-5}$ and $(16.0 \pm 5.6) \times 10^{-5}~\text{MeV cm}^{-2} \text{ s}^{-1}$, in the pertinent time bins. However, given the proximity of this source to the galactic plane, the signal could be caused by contamination from diffuse emission. Additionally, the presence of the X-ray binary 2E-2336 within the 99\% uncertainty region suggests that it could be a potential contributor to the detected gamma-ray signal.

    
    Therefore, except for one object, we can definitely rule out gamma-ray emission in coincidence with the magnetar X-ray outbursts with our analysis. As discussed in ~\cite{Abdo10} and ~\cite{Li17}, the non-detections rule out most of the parameter space of the outer gap model for magnetars proposed in ~\cite{Cheng01}. Our flux limits for 1E2259+586 are of the same order and magnitude as ~\cite{Li17} and also rule out the model for GeV gamma-ray emission proposed in ~\cite{Takata13}.

    In a forthcoming work, we shall extend this search to extragalactic magnetars around the time of giant flares, such as GRB 200415A and GRB 231115A.

\begin{acknowledgements}
The authors thank Siddhant Manna for useful discussions and Raniere Menezes for help with {\tt easyFermi}. We are also grateful to the anonymous referees for very useful feedback on the manuscript.
\end{acknowledgements}

\bibliography{main}

\end{document}